# The Promise of Low-Frequency Gravitational Wave Astronomy

Lead Author: Tom Prince (Caltech/JPL)
for
Members of the LISA International Science Team

Peter Bender (U. of Colorado)
Sasha Buchman (Stanford)
Curt Cutler (JPL)
Joan Centrella (GSFC)
Sam Finn (Pennsylvania State University)
Jens Gundlach (U. of Washington)
Craig Hogan (U. of Chicago/Fermilab)
Scott Hughes (MIT)
Piero Madau (U. of California, Santa Cruz)
Sterl Phinney (Caltech)
Douglas Richstone (U. of Michigan)
Robin Stebbins (GSFC)
Kip Thorne (Caltech)

Pierre Binetruy (APC, College de France)
Massimo Cerdonio (U. of Padova)
Mike Cruise (U. of Birmingham)
Karsten Danzman (U. of Hannover/MPI)
James Hough (Glasgow University)
Oliver Jennrich (ESTEC)
Philippe Jetzer (U. of Zurich)
Alberto Lobo (U. de Barcelona)
Yannick Mellier (Institut d'Astro. de Paris)
Bernard Schutz (AEI Potsdam/MPI)
Tim Sumner (Imperial College)
Jean-Yves Vinet (Obs. de la Côte d'Azur)
Stefano Vitale (U. of Trento)

**Relevant Astro2010 Science Frontier Panels:**
Galaxies across Cosmic Time
Stars and Stellar Evolution
The Galactic Neighborhood
Cosmology and Fundamental Physics

*[This science white paper provides an overview of the opportunities in low-frequency gravitational-wave astronomy, a new field that is poised to make significant advances in coming years. While discussing the broad context of gravitational-wave astronomy, this paper will concentrate on the low-frequency region ($10^{-5}$ to 1 Hz), a frequency range abundantly populated in strong sources of gravitational waves. This paper is complementary to other science white papers being submitted that will discuss specific sources and science objectives in more detail. ]*

# 1. The Emerging Field of Gravitational Wave Astronomy

Gravitational-wave (GW) astronomy is poised to make revolutionary contributions to astronomy and physics during the next two decades. In particular, the low frequency GW region ($10^{-5}$ to 1 Hz) is rich in guaranteed sources of strong gravitational waves and offers tremendous potential for new and unexpected discoveries.

Historically the most important advances in astronomy have been driven by major leaps in observational capability. Consider a new observational window for astronomy that opens up 4-5 decades of frequency for study, that is capable of observing sources with high signal-to-noise out to very high redshift ($z > 15$), and that can provide measurements of important astrophysical quantities (mass, spin, orbits, luminosity distance) to percent-level accuracy. If this were a window of the electromagnetic spectrum, the arguments for opening it would be absolutely compelling. That this window is also non-electromagnetic makes it a radically new capability that provides an entirely new dimension to observational astronomy.

To date most of our information about the universe comes from electromagnetic observations, with important contributions also coming from astroparticle measurements. Until we observe gravitational waves, however, our picture of the universe and its constituents will be incomplete. Objects of fundamental importance such as astrophysical black holes merge and radiate with luminosity larger than the entire electromagnetic universe, yet these events remain largely invisible without GW observations. Gravitational wave astronomy will provide us with a clear, detailed, and extinction-free view of the astrophysical sources and processes driven by strong gravitational fields.

When a radically new capability is realized, the opportunity for discovery is very high. Given the history of astronomy, it would be astounding if opening 4-5 orders of magnitude of non-electromagnetic frequency space with high sensitivity did not yield paradigm-changing results. When observed with GWs, intrinsically interesting astronomical sources such as massive black holes in galactic nuclei and ultra-compact stellar binaries will surely yield many new surprises. The discovery potential is immense.

Low-frequency GW astronomy is particularly promising. The wavelengths and frequencies covered ($\sim 10^5$ to $10^{10}$ km and $\sim 10^{-5}$ to 1 Hz) correspond to very interesting astrophysical scales, in particular the orbital size and temporal scales for both massive black hole binaries and compact stellar-mass binaries (e.g. double white dwarf binaries). The strong, coherent nature of the GW signals and their time dependence yield highly precise information about the radiating systems, on size scales difficult to probe using conventional approaches. In addition, the strong temporal evolution of the orbits of many GW sources provides a further important capability: the ability to measure intrinsically precise, systematic-free luminosity distances [e.g. Schutz 1986; Lang & Hughes 2008].

Low-frequency gravitational wave (GW) astronomy should be seen within the context of the broader field of gravitational waves, which spans a wavelength range of over 18 orders of magnitude and which encompasses a tremendous range of physics and astrophysics.

Figure 1, from the NASA Beyond Einstein Program, shows the immense span of GW measurements.

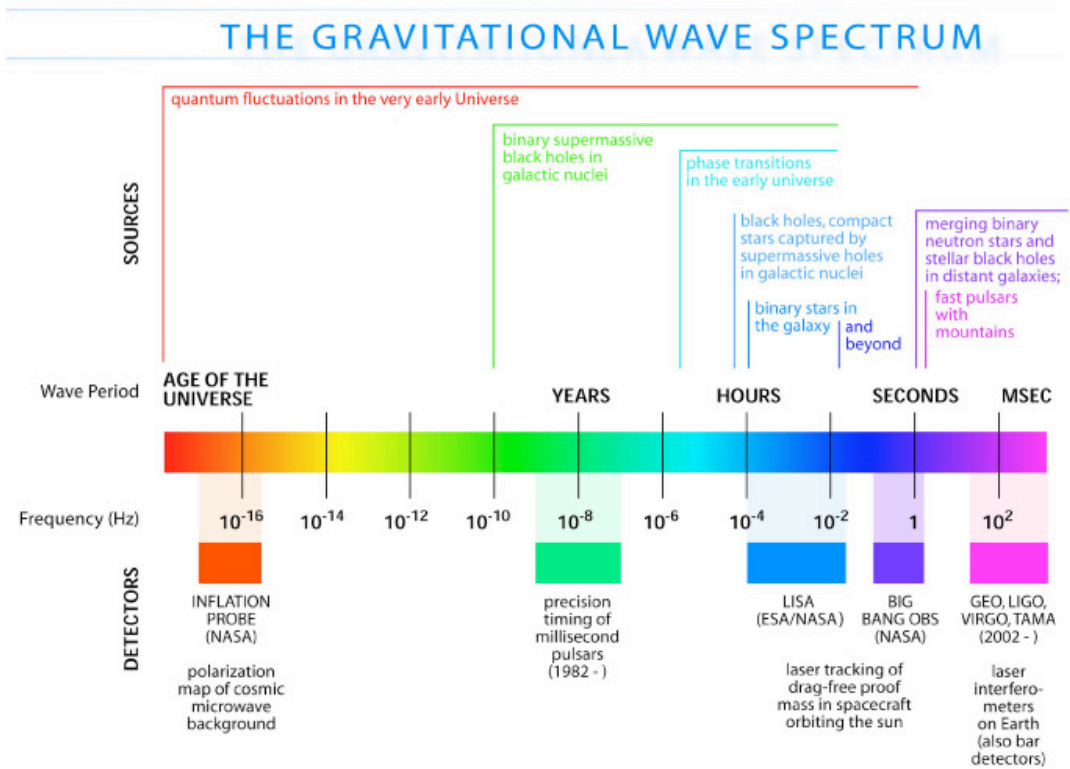

**Figure 1**

The figure indicates that GW observations will be undertaken in several different frequency regimes, using a variety of approaches:

- **CMB:** Indirect observations of the imprint of GW on the early universe
- **VLF:** Very-low-frequency observations of the diffuse background from massive black hole mergers via pulsar timing (see, e.g. NANOGrav 2009)
- **LF:** Low-frequency ($10^{-5}$ – 1 Hz) observations from space (LISA and follow-on missions)
- **HF:** High-frequency (10 Hz – $10^4$ Hz) observations from the ground (LIGO, Virgo, etc.)

The field of gravitational-wave astronomy has the potential to address many of the most fundamental questions of astronomy and physics including (with an indication of the relevant frequency range):

- How did galaxies and black holes co-evolve over the history of the Universe? (VLF, LF)
- What are the dynamics of stars in the dense environment of galactic nuclei? (LF)
- What are the extreme endpoints of binary stellar evolution? (LF, HF)
- What is the physics of gravitational collapse in supernovae? (HF)

- What are the physics and astrophysics of matter in extreme conditions? (LF, HF)
- What are the limits of General Relativity as a description of gravity? (LF, HF)
- What is the physics of inflation? (CMB, LF, HF)
- What is the nature of dark energy and the equation of state of the universe? (LF)
- Is there astrophysical evidence for physics beyond the standard model? (VLF, LF, HF)

GW observations complement electromagnetic observations in providing answers to these and other important questions.

- JWST, ALMA, and other facilities will probe the evolution and merger of galaxies at high redshift; GW observations will provide the corresponding picture for the co-evolution and merger of massive black holes (MBHs) at high redshift.
- X-ray observations tell us how black holes grow by accretion; GW observations will tell us how they grow by mergers.
- Electromagnetic observations across the spectrum provide information on black holes in active galaxies; GW can provide us with highly accurate measurements of the masses and spins of quiescent black holes in normal galaxies.
- Electromagnetic observations provide information on the disruption of stars by the massive black holes in the nuclei of galaxies; GW observations will give us the complementary information about the capture of compact objects (BH, NS, WD) by massive black holes, which are mostly invisible to EM observations.
- Electromagnetic observations of stellar-mass compact binaries provide us with a wealth of information, in particular through their mass-transfer signature; GW measurements provide information about the heaviest and most compact binaries, whether or not they are transferring mass, and will increase our census of these objects by a factor of ~100.

A particularly exciting possibility is the observation of electromagnetic counterparts to GW sources. A detailed discussion of observing such counterparts is beyond the scope of this white paper, but is discussed in several other Astro2010 science white papers [e.g. Bloom et al. 2009; Phinney 2009]. The scientific opportunities presented by combining EM and GW observations are many and greatly enhance the potential science from observations of massive black holes and ultra-compact stellar binaries, as well as open up additional possibilities such as precision measurement of dark energy and cosmological parameters.

## 2. Opportunities in Low-Frequency Gravitational Wave Astronomy

Low-frequency GW observations will impact many areas of astronomy and physics. We briefly mention several specific opportunities here. Additional details can be found in other Astro2010 science white papers referenced below. Areas of opportunity include:

- *Determination of the history of massive BH mergers in the universe,* including precision measurements of masses, mass ratios, spins, and luminosity distances out to very large distance ($z\sim 20$ if mergers occur at that distance). Such information will dramatically impact our picture of galaxy evolution and the role of BHs in that evolution [e.g. Madau et al. 2009].
- *Determination of the nature of the "seed black holes" from which the massive black holes in today's galaxies evolved.* Characterizing the mass and spin of the BHs in the earliest mergers will provide significant constraints on the nature of the seed BHs [e.g. Sesana et al. 2008].
- *Determination of the population and dynamics of compact stellar objects (BH, NS, WD) within the sphere of influence of the massive BHs in the centers of galaxies.* Observations of inspiral and capture of compact objects by massive BHs will provide a wealth of information (MBH mass and spin, compact object mass and orbit) of a population of objects certain to exist in the nuclei of galaxies, but unobservable in all galaxies but perhaps our own [e.g. Amaro-Seoane et al. 2007; Miller et al. 2009].
- *Determination of the population of ultra-compact binaries in our galaxy.* This will yield new insights into the evolutionary channels that produce such binaries as well as the physics of the interactions of the binary components. The first generation of instruments will be capable of individually identifying and characterizing ten thousand or more ultra-compact binaries, both accreting and non-accreting (mostly WD-WD binaries) [e.g. Nelemans 2008 & 2009]. Follow-on missions to LISA in the deciHz range offer the exciting prospect of individually identifying and characterizing ultra-compact NS-NS, NS-BH, and BH-BH binaries in the universe out to very high redshift [e.g. Cutler and Harms 2006].
- *Tests of General Relativity with high precision.* These will be made in both the highly dynamical strong-field regime using massive BH mergers [e.g. Berti et al. 2005], and in the "test-mass" regime using observations of the hundreds of thousands of orbits traversed as a stellar-mass compact object (WD, NS, or BH) spirals into a MBH [e.g. Glampedakis & Babak 2006; Barack & Cutler 2007; Schutz et al. 2009].
- *Independent and precise measurements of cosmological parameters.* GW observations can yield measurements of the luminosity distance with an intrinsic precision of better than 1% out to z~1, with the ultimate accuracy limited by weak lensing to ~3%. The determination of the luminosity distance is very clean, depending only on General Relativity. If an EM counterpart is identified for a massive BH merger, determination of the redshift of the host galaxy immediately yields a measurement of the distance-redshift relationship [e.g. Holz & Hughes 2005; Kocsis et al. 2006; Hogan et al. 2009a]. This provides the prospect of an independent determination of dark-energy parameters. It has also been proposed that GW luminosity-distance measurements for the capture of compact objects by MBH can yield an estimate of the Hubble constant to 1% accuracy [MacLeod & Hogan 2008], even in cases where the host galaxies are not identified.
- *Prospects for the direct observation of radiation from the post-inflationary era of the universe* ($\sim 10^{-18}$-$10^{-11}$ seconds after the Big Bang). GWs probe length and energy scales 14 orders of magnitude shorter and more energetic than CMB observations. First generation low-frequency GW observations have the possibility of directly observing GWs from phase transitions in the early universe [e.g. Grojean & Servant 2007; Hogan et al. 2009b]. Future missions may be able to reach sensitivities needed to observe the levels predicted by standard slow-roll GUT-scale inflation [e.g. Boyle & Steinhardt 2008].

- *Potential for new discoveries.* Low-frequency GW observations could yield the first evidence for cosmic strings, evidence for deviations from the standard model, evidence for extra dimensions, or for any number of other possible discoveries.

Many of the opportunities mentioned above will be realized by observing particular classes of low-frequency GW sources, summarized in Table 1. Additional details of sources are given in other Astro2010 white papers (see references). The Laser Interferometer Space Antenna (LISA) will study each of these classes of sources with high sensitivity (see the additional sources of information provided in section 4 of this white paper).

| Table 1: Sources for Low-Frequency GW Astronomy ($10^{-5} - 1$ Hz) ||
|---|---|
| Massive Black Hole (MBH) Merger ||
| Characteristics | Inspiral and merger of MBH binary ($z = 0 - 20$) |
|  | Mass Range: $10^4 - 10^7\ M_\odot$; Orbital Period: $10^3 - 10^5$ s |
| Observables | Masses: $\frac{\sigma_M}{M} \lesssim 1\%$; Spins: $\frac{\sigma_S}{S} \lesssim 1\%$ |
|  | Luminosity Distance: $\frac{\sigma_{D_L}}{D_L} \lesssim 3\%$ @ $z = 1$ (limited by weak lensing) |
| Science Objectives | • Nature of seed BHs |
|  | • Growth and evolution of BHs |
|  | • Tests of General Relativity in strong-field, highly dynamical regime |
| Capture of Stellar Mass Compact Objects by MBH ||
| Characteristics | Compact-object (BH, NS, or WD) inspirals into massive BH |
|  | MBH Mass: $10^4 - 10^7\ M_\odot$; Orbital Period: $10^3 - 10^5$ s |
| Observables | Masses: $\frac{\sigma_M}{M} \lesssim 0.01\%$; Spins: $\frac{\sigma_S}{S} \lesssim 0.1\%$ |
|  | Luminosity Distance: $\frac{\sigma_{D_L}}{D_L} \lesssim 1\%$ @ z=0.5 (limited by weak lensing) |
| Science Objectives | • Masses and spins of nuclear BHs in normal galaxies |
|  | • Population and dynamics of compact objects in galactic nuclei |
|  | • Precision tests of General Relativity |
| Ultra-Compact Binaries ||
| Characteristics | Primarily compact WD-WD binaries; mass transferring or detached |
|  | $\geq 10{,}000$ individual sources + diffuse background at low frequencies |
|  | Orbital periods: $\sim 10^2 - 10^4$ s |
| Observables | Orbital frequency; Chirp mass; Frequency derivative; Distances |
| Science Objectives | • Evolutionary pathways, e.g. outcome of common envelope evolution |
|  | • WD-WD as possible SN1a progenitors |
|  | • Tidal interactions and mass transfer |

## 3. Evolution of Science and Technical Capabilities

The science that can be addressed using low-frequency gravitational waves continues to grow more exciting as time goes on, and there has been a significant evolution in the scope and impact of that science since the last decadal review. Among the many relevant advances since the last review are:

- The co-evolution of galaxies and their massive black holes has become a major focus of astronomical research. GW observations will directly detect the often-invisible MBH and their binary coalescence, providing critical new information on distances and rates as well as precision measurements of the mass and spin of binary constituents. These observations will complement the observations of high-redshift galaxies from JWST, ALMA, and other ground- and space-based observatories.
- Interest has grown in determining the nature of the seed black holes that grew into the massive black holes in the centers of galaxies. Did they arise from PopIII stars, collapse of dense early star clusters, or some other mechanism? GW observations of the earliest mergers at very high redshift will be critical.
- The beautiful results from high-spatial-resolution measurements of our Galactic Center (with Keck, ESO), from detections of hyper-velocity stars in our Galaxy, and from observations of stellar disruptions in external galaxies (e.g. ROSAT, Galex) all indicate that the nuclei of galaxies are fascinating dynamical environments with much still to be learned about them. GW observations will provide unique and detailed observations of the capture of stellar-mass compact objects by MBHs in galactic nuclei.
- The study of Dark Energy has benefited from new and innovative observational approaches. As recognized in the recent NRC review of the NASA Beyond Einstein program, GW measurements have the potential to contribute uniquely to determining the distance scale of the Universe through precise determination of absolute distances.

Likewise, there has been significant evolution in the technical capabilities needed for low-frequency GW astronomy:

- Recent breakthrough progress in numerical General Relativity has provided a strong foundation for interpreting the precision measurements from both ground-based and space-based gravitational wave observatories [e.g. Baker et al. 2007].
- The GW ground-based, space-based, and pulsar-timing communities have developed powerful analysis capabilities. For low-frequency GW observations, we note in particular the success of the Mock LISA Data Challenges in demonstrating source parameter determination using realistic simulated data [e.g. Arnaud et al. 2007], as well as the ability to simultaneously identify ~20,000 WD binaries.
- Significant instrumental progress has been realized. While not the subject of a science white paper, we note that the instrumental technologies required for low-frequency GW astronomy are now at a mature stage, with numerous laboratory demonstrations (e.g. via torsion balance and laser interferometry measurements — see e.g. http://www.srl.caltech.edu/lisa/mission_documents.html) and an upcoming technology validation mission, LISA Pathfinder [McNamara 2008]. The LISA mission architecture for low-frequency GW astronomy is well-studied and mature.

As a consequence of these advances in science and technology over the last decade, low-frequency GW astronomy is poised to realize its scientific potential. It offers a wealth of new scientific opportunities. Low-frequency GW astronomy will provide unique observations of familiar sources, open a new window on entire classes of previously invisible sources, and offer immense potential for unexpected discoveries.

## 4. References and Additional Information

Additional information about low-frequency gravitational waves astronomy, primarily in the context of the LISA mission, includes:

- LISA: Probing the Universe with Gravitational Waves (http://www.srl.caltech.edu/lisa/documents/lisa_science_case.pdf)
- Overview papers on low-frequency gravitational wave science and technology can be found at http://www.srl.caltech.edu/lisa/mission_documents.html and at http://www.lisascience.org/
- LISA science is well summarized in the NRC Beyond Einstein Assessment Committee report http://books.nap.edu/catalog.php?record_id=12006